\overfullrule=0pt   \magnification=\magstep1  \input amssym.def
\font\huge=cmr10 scaled \magstep2 
\voffset.4in
\def\i{{\rm i}}  \def\si{\sigma}   \def\eps{\epsilon} 
\def\Z{{\Bbb Z}}  \def\Q{{\Bbb Q}}   

\def\boxit#1{\vbox{\hrule\hbox{\vrule{#1}\vrule}\hrule}}
\def\stimes{\,\,{\boxit{$\times$}}\,\,}  \def\om{{\omega}}
\font\smal=cmr7
\font\smit=cmmi7
 

\centerline{{\huge Integers in the Open String}}\medskip
\bigskip 

\centerline{{Terry Gannon}}
\medskip\centerline{{\it Department of Mathematical Sciences,}}
\centerline{{\it University of Alberta}}
\centerline{{\it Edmonton, Canada, T6G 2G1}}\smallskip
\centerline{{tgannon@math.ualberta.ca}}  \bigskip\medskip


{\smal We show that the {\smit Y}$_{{\smal ab}}^{\,{\smal c}}$ of Pradisi-Sagnotti-Stanev are indeed
integers, and we prove a conjecture of Borisov-Halpern-Schweigert. We
indicate some of the special features which arise when the order of
the modular matrix {\smit T} is odd. Our
arguments are general, applying to arbitrary ``parent'' RCFT assuming
only  that {\smit T} has odd order.}\bigskip\bigskip

\noindent{{\bf 1.\ Introduction}}\medskip

There has been considerable interest recently in analysing aspects of
the open string in terms of the
``parent'' rational conformal field theory (RCFT) on 2-dimensional surfaces with boundaries (this strategy is
called the method of open descendants) --- 
see e.g.\ [1--6] and references therein for samples from
this work.

Cardy's influential work [7] investigated the boundary conditions on a
cylinder (=annulus), for those RCFT whose torus partition function is
given by the charge-conjugation matrix $C=S^2$: 
$${\cal Z}(\tau)=\sum_a\chi_a\,\chi_{\overline{a}}^*\ .\eqno(1)$$
In this case, the boundary conditions and primary fields are in
one-to-one correspondence, and the annulus coefficients $A_{ab}^c$ are
given by the chiral fusion coefficients $N_{ab}^c$. In other words,
Cardy discovered a new interpretation for the coefficients $N_{ab}^c$: they also
count the number of open string states in sector $c$ with boundary
conditions $a$ and $b$.

Specifying the theory also requires knowing the M\"obius strip
and Klein bottle coefficients
$M_a^b$ and $K_a$. These coefficients must satisfy certain consistency
conditions, and Pradisi-Sagnotti-Stanev [8] expressed a solution in
terms of an ``integer-valued tensor'' $Y_{ab}^c$ (see (9) below). These numbers
$Y_{ab}^c$ also appear in the fusion coefficients of ${\Bbb
Z}_2$ permutation-orbifolds [9,10].

In this paper we find a novel interpretation for these $Y_{ab}^c$
(up to a sign they equal certain fusion coefficients!) which makes it
manifest  that they indeed are integers. A general proof
of this seems to be lacking in the literature (see the last paragraph
of this section). This also yields a
positivity condition on the $K_a$. We also use this to prove a
conjecture in [9].

Our results are very general: they apply to {\it any} RCFT, provided
only that the order of the modular matrix $T={\rm
diag}(\exp[2\pi\i\,(h_a-{c\over 24})])$ is {\it odd}. 
In particular, we require that Verlinde's formula for the ``parent''
RCFT yields nonnegative integers, and that equations (5) below hold.
The odd-order hypothesis 
is vital for our interpretation, and for the positivity result on the
$K_a$, but both the integrality of the $Y_{ab}^c$ and the conjecture of
[9] should continue to hold when $T$ has even order.

In a recent paper [11], Bantay also relates the conjecture of [9] to
that of [8]. In particular, he proves the integrality of $Y$, assuming
that Verlinde's formula will yield nonnegative integers for an
appropriate cyclic orbifold, and assuming results from the theory of
permutation orbifolds [9]. On the one hand, his argument does not
assume $T$ has odd order. On the other hand, his argument is built on
a somewhat less solid foundation than that of this paper: e.g.\ the
formula for the modular matrix $S$ for cyclic orbifolds is read off
from expressions for the orbifold characters, but that technique works
only when  those
characters are linearly independent, which for general RCFT is false.
More precisely, the formula (2a) below  uniquely specifies the
coefficients $S_{ab}$ only if the characters ${\rm ch}_b(\tau)$ are
linearly independent, and yet e.g.\ ch$_{\overline{a}}(\tau)={\rm ch}_a(\tau)$.
Reference [9] includes Cartan angles (corresponding to commuting spin
one currents) in their characters, but there seems to be no reason why
these should be sufficient to guarantee linear independence for
arbitrary  RCFT. It thus appears that the
formulas [9,10] for the modular matrix $S$ in permutation orbifolds is conjectural.

\bigskip\noindent{{\bf 2.\ Background material}}\medskip

The characters $\chi_a(\tau)$ of an RCFT carry a representation of
SL$_2(\Z)$ [12]:
$$\eqalignno{\chi_a(-1/\tau)=&\,\sum_bS_{ab}\,\chi_b(\tau)&(2a)\cr 
\chi_a(\tau+1)=&\,\sum_bT_{ab}\,\chi_b(\tau)&(2b)} $$
The subscripts $a,b$ label the (finitely many) primary fields.
The matrices $S$ and $T$ are unitary and symmetric, and $T$ is
diagonal with entries exp$[2\pi\i\,(h_a-{c\over 24})]$ where $h_a$ are the
conformal weights and $c$ is the central charge. $T$ has finite order [13]:
i.e.\  $T^N=I$ for some $N$. The matrix
$C=S^2$ is the order-2 permutation matrix called charge-conjugation;
we'll write $\overline{a}$ for $Ca$.

There are two choices of $2\times 2$ matrices corresponding to the
fractional linear transformation $\tau\mapsto -1/\tau$ of (2a); we'll
adopt the more common choice in RCFT, namely  $\left(\matrix{0&-1\cr
1&0}\right)$. For this choice, $(ST)^3=(TS)^3=C$.

The fusion coefficients $N_{ab}^c$ can be expressed using
Verlinde's formula:
$$N_{ab}^c=\sum_d{S_{ad}\,S_{bd}\,S^*_{cd}\over S_{0d}}\eqno(3)$$
where `0' denotes the vacuum.

We will need two lesser known properties of RCFT. Both hold for any
RCFT.  The first is due to Bantay [14].
Define the numbers
$$Z(a,b):=\exp[-\pi\i\,h_b]\sum_{i,j}N_{ij}^a\,S_{bi}^*\,S_{0j}\,T^2_{jj}T^{-2}_{ii}\eqno(4)$$
Then 
$$\eqalignno{Z(a,b)\equiv&\,N_{aa}^b\quad({\rm mod}\ 2)&(5a)\cr
|Z(a,b)|\le &\,N_{aa}^b&(5b)}$$
$Z(a,0)$ is Bantay's Frobenius-Schur indicator: $Z(a,0)=0$ if $a\ne
\overline{a}$, and $Z(a,0)=\pm 1$ depending on whether $a$ is
real or pseudo-real.

The other property is the Galois symmetry [15]. The entries of $S$
will lie in some cyclotomic extension of $\Q$ --- that just means each
$S_{ab}$ can be written as a polynomial $p_{ab}(\xi_n)$, with rational coefficients,
evaluated at some root of unity $\xi_n:=\exp[2\pi\i/n]$. By the ``Galois
group'' of this cyclotomic extension, we mean the integers mod $n$
which are coprime to $n$ --- we write this $\Z_n^\times$. For example,
$Z_{12}^{\times}=\{1,5,7,11\}$. Any such
$\ell\in\Z_n^\times$ acts on the number $p_{ab}(\xi_n)$ by sending it to
$p_{ab}(\xi_n^\ell)$ --- we denote  this ``Galois automorphism'' by $\si_\ell$.
For instance, $\si_\ell$ fixes all rationals, $\sigma_\ell\,\cos(2\pi{a\over n})=\cos(2\pi{\ell a\over n})$,
$\si_\ell \,\i=\pm 1$ when $\ell\equiv \pm 1$ (mod 4),
and $\si_\ell\sqrt{2}=\pm\sqrt{2}$ depending on whether or not $\ell\equiv \pm 1$
(mod 8).

The point [15] is that to any $\ell\in\Z_n^\times$, there is a permutation
$a\mapsto \si_\ell a$ of the primary fields, and a choice of signs
$\epsilon_\ell(a)=\pm 1$, such that 
$$\si_\ell(S_{ab})=\epsilon_\ell(a)\,S_{\si_\ell
a,b}=\epsilon_\ell(b)\,S_{a,\si_\ell b}\eqno(6a)$$
See \S4 for some examples. Define the signed permutation matrix $G_\ell$ by
$(G_\ell)_{ab}=\epsilon_\ell(a)\,\delta_{b,\si_\ell a}$. Then (6a) can
be written as
$$\si_\ell\,S=G_\ell S=SG_{\ell}^{-1}\ .\eqno(6b)$$
These orthogonal matrices $G_\ell$ in fact constitute a representation of the multiplicative
group $\Z_n^\times$, i.e.\ $G_\ell\,G_m=G_{\ell m}=G_m\,G_\ell$.

This Galois symmetry should be regarded as a generalisation of charge-conjugation.
In particular, $\ell=-1$ corresponds to the rather familiar Galois automorphism $\si_{-1}=*$
(complex conjugation), and $\eps_{-1}(a)=+1$, $\si_{-1}a=\overline{a}$,
and $G_{-1}=C$. The commuting of the $G_\ell$ then implies $\eps_\ell(\overline{a})
=\eps_\ell(a)$ and $\si_\ell(\overline{a})=\overline{\si_\ell(a)}$. 

Next, turn to the open string amplitudes. The case we will consider
here is the best-understood one: where the torus partition function is
given by (1).

There are two channels (transverse and direct) for the annulus, Klein
bottle  and M\"obius strip, corresponding to the two different choices
of time (``horizontal'' and ``vertical''). The two channels give
identical amplitudes, provided the modular transformation relating the
natural modular parameters in the two channels is taken into
consideration. This modular transformation is $S$, for both the annulus and
Klein bottle, but is
$$P:=\sqrt{T}\,S\,T^2\,S\,\sqrt{T}\eqno(7)$$
for the M\"obius strip. By `$\sqrt{T}$' here we mean the diagonal matrix
with entries $\exp[\pi\i\,(h_a-{c\over 24})]$. Note that $P$ is
unitary and symmetric and $P^2=C$, hence $P^*=CP=PC$.

The direct channel
amplitude for the annulus is the open string partition function: $$A_{ab}={1\over 2}\sum_cA_{ab}^c\,\chi_c\eqno(8a)$$
where as already mentioned $A_{ab}^c=N_{ab}^c$ here. The direct
channel amplitudes for the M\"obius strip and Klein bottle are
$$\eqalignno{M_a=&\,\pm{1\over 2}\sum_bM^b_a\,\hat{\chi}_b&(8b)\cr
K=&\,\sum_a K^a\,\chi_a&(8c)}$$
where in (8b) the $\hat{\chi}_a$ are a basis of ``real characters''
(see [1,2] for details).

There are a number of constraints on these coefficients [1,2], and the task
is to find solutions to them. Cardy
gives $A_{ab}^c=N_{ab}^c$, and Pradisi-Sagnotti-Stanev give
$M_a^b=Y_{a0}^b$ and $K^a=Y_{\overline{a}0}^0$, where $Y_{ab}^c$ is given by
$$Y_{ab}^c:=\sum_{d}{S_{ad}\,P_{bd}\,P_{cd}^*\over S_{0d}}\eqno(9)$$
Note that $Y_{0b}^c=\delta_{b,c}$ and $Y_{ab}^c=Y_{\overline{a}c}^b=Y_{\overline{a}
\overline{b}}^{\overline{c}}$; from the given properties of $P$, $Y_{ab}^c$
will automatically be real.
Writing $Y_a$ for the matrix with entries $Y_{ab}^c$, we find that
$Y_a$ give a representation of the fusion ring:
$Y_aY_b=\sum_cN_{ab}^cY_c$. 

It is conjectured in [8] that the $Y_{ab}^c$ are integers (though not
necessarily positive). There does not appear to be a proof of this, at
least not in full generality (see the last paragraph of \S1), although [16] show that
$Y_{a0}^b=Z(a,b)$ and hence those particular numbers are necessarily integers
by (5a). In the next section we demonstrate the integrality of all $Y_{ab}^c$,
 assuming only that the order $N$ of $T$ is odd.

There is no unique solution to the various constraints: e.g.\ [16]
shows that provided simple-currents are present, there will be
others. We will consider here only the ``standard'' Klein bottle of
[8], given above.

\bigskip\noindent{{\bf 3.\ The open string when $N$ is odd}}\medskip

This section is the heart of the paper.

For the remainder of this paper, assume the order $N$ of $T$ is odd.
This assumption can be rephrased as follows. Write $t(r)$ for the exponent of
2 that appears in the
prime decomposition of the rational number $r$: e.g. $t(3.7)=-1$.
Then $N$ odd means that each $t(h_a)\ge 0$, as well as $t(c)\ge 3$.

Examples of odd $N$ occur for instance in the WZW theories $SU(M)$
level $k$ when both $M$ is odd and $k$ is even. Another large class
of examples are holomorphic orbifolds (with or without discrete torsion)
by a finite group $G$ with odd order.

The main result of this paper is the following new expression for
$Y_{ab}^c$:
$$Y_{ab}^c=s(b)\,s(c)\,\epsilon_{{1\over 2}}(b)\,\epsilon_{{1\over
2}}(c)\,N_{a,\si b}^{\si c}\eqno(10a)$$
where `$\si$' denotes the Galois permutation $\si_{{1\over 2}}$, and
where $s(a):=+1$ if $t({h_a}-{c\over 24})> 0$, otherwise
$s(a):=-1$. The fraction `${1\over 2}$' here denotes the mod $N$ inverse of 2, i.e.\
${1\over 2}={N+1\over 2}\in\Z_N^\times$. So {\it up to a sign, $Y_{ab}^c$
is a fusion coefficient!} Similarly, we obtain the
curious expression
$$P=s\,S\,G_2\,s=\si_{{1\over 2}}(s\,S\,s)\eqno(10b)$$
where $s={\rm diag}(s(a))$ and the matrix $G_2$ is as in (6b). We will prove
equations (10) in the final section. But first
let's explore some of their consequences.

Since the $\epsilon$'s and $s$'s in (10a) are signs, and fusion coefficients are
nonnegative integers, {\it the $Y_{ab}^c$ will necessarily be integers}.

Recall the observation [16] that $Y_{a0}^b=Z(a,b)$. Then (5)
become [17] the unobvious
$$\eqalignno{N_{a,\si 0}^{\si b}\equiv &\,N_{aa}^b\quad({\rm mod}\
2)&(11a)\cr N_{a,\si 0}^{\si b}\le &\,N_{aa}^b\ .&(11b)}$$

Putting $b=0$ in (11) gives a curious fact [17]: the fusion product of
$\si_{{1\over 2}}(0)$ with itself consists of all self-conjugate fields,
each appearing with multiplicity 1!

The M\"obius strip and Klein bottle coefficients become
$$\eqalignno{M_a^b=&\,\epsilon_{{1\over 2}}(b)\,\epsilon_{{1\over 2}}(0)\,
N_{a,\si b}^{\si 0}&(12a)\cr
K^a=&\,N_{\si 0,\si 0}^a&(12b)}$$
(12b) is the interesting one: most curiously, it is never negative!
In particular, it is now a consequence of (11) that
$$K^a=\left\{\matrix{1&{\rm if}\ a=\overline{a}\cr 0&{\rm
otherwise}}\right.\eqno(12c)$$

This positivity is related to an observation made in [17]: when $N$
is odd, there are no pseudo-real primary fields. The signs
$\epsilon_a$ of the coefficients $K^a$ play a role in the open string
theory [1,2], and when $N$ is odd we see they all equal +1.

As a final result, let us prove a conjecture made in [9]. In their
equation (4.38), they find that the combination
$${1\over 2}\sum_d{S^2_{ad}\,S_{bd}S_{cd}^*\over S^2_{0d}}\pm{1\over
2}Y_{ab}^c\eqno(13)$$
should equal a fusion coefficient in a $\Z_2$ cyclic-orbifold theory. They
conjecture that for any RCFT, and any choice of signs, (13) should be
a nonnegative integer.

First, note that the first term (dropping the factor ${1\over 2}$) is
$$(N_aN_a)_{bc}=\sum_d N_{aa}^d\,N_{db}^c\ ,\eqno(14a)$$
since the {\it fusion matrices} $N_a$ (with entries
$(N_a)_{bc}:=N_{ab}^c$) form a representation of the fusion ring.
Using (10a), the second term can be written as
$$N_{a,\si b}^{\si c}=\sum_d{\epsilon_{{1\over
2}}(d)\,\epsilon_{{1\over 2}}(0)\over \epsilon_{{1\over
2}}(b)\,\epsilon_{{1\over 2}}(c)}\,N_{bd}^c\,N_{a,\si 0}^{\si d}\ ,\eqno(14b)$$
where we use equation (18) of [15].
In order to prove the conjecture of [9], we must show both that (14a) and
(14b) are congruent mod 2,
and that the absolute value of (14b) is bounded above by (14a).
Their congruence mod 2 follows immediately from (11a), and the desired   
inequality follows from (11b) and the triangle
inequality applied to (14b).

Thus, provided the order of $T$ is odd, we have proven the conjecture
of [9] that (13) is always a nonnegative integer.

The assumption that the order $N$ of $T$ is odd, is vital for the
interpretation involving the Galois symmetry. For further 
consequences of $N$ being odd, see \S\S2.5,3.3 of [17].

It would be very interesting to find interpretations \`a la (10) for
the other integral representations $Y_a^{(\ell)}:=G_\ell^{-1}N_a
G_\ell$ of the fusion ring. This should involve the ``twisted
dimensions'' ${\cal D}_0\left({p\ q\
r\atop 0\ {1\over \ell}\ {1\over \ell}}\right)$ of [11] --- see also \S3.3 of [17].
\bigskip

\noindent{{\bf 4.\ Examples}}\medskip

In this section we work out the relevant quantities for two WZW examples.
The WZW model for $SU(M)$ level $k$ has primaries most conveniently
denoted by $M$-tuples of nonnegative integers which sum to $k$.

Consider first the WZW model for $SU(3)$ at level 2. There are 6
primaries: (2,0,0) (the vacuum `0'), (0,2,0) and
(0,0,2) (the two nontrivial simple-currents), and (1,1,0), (1,0,1) and (0,1,1).
The matrix $T$ is diag$(\xi^{-2},\xi^8,\xi^8, 
\xi^2,\xi^2,\xi^7)$ where $\xi=\exp[2\pi\i/15]$, so its order is
$N=15$. The matrix $S$ here equals
$$S={2\,\sin({\pi\over 5})\over \sqrt{15}}\left(\matrix{1&1&1&2c&2c&2c\cr
1&\om^2&\om&2\om c&2\om^2 c&2c\cr 1&\om&\om^2&2\om^2 c&2\om c&2c\cr
2c&2\om c&2\om^2c&-\om^2&-\om&-1\cr 2c&2\om^2c&2\om c&-\om&-\om^2&-1\cr
2c&2c&2c&-1&-1&-1\cr}\right)$$
where $\om=\exp[2\pi\i/3]$ and $c=\cos({\pi\over 5})$.
The matrix $s$ used in equations (10) is $s={\rm diag}(+1,+1,+1,+1,+1,-1)$.
Write $\si$ for $\si_{{1\over 2}}$; then $\si\sqrt{15}=-\sqrt{15}$,
$\si\sin({\pi\over 5})=\sin({2\pi\over 5})$, $\si\om=\om^2$, and $\si\cos({
\pi\over 5})=\cos({2\pi\over 5})$. The Galois matrix is thus
$$G_{{1\over 2}}=G_2^t=\left(\matrix{0&0&0&0&0&-1\cr 0&0&0&-1&0&0\cr
0&0&0&0&-1&0\cr 0&-1&0&0&0&0\cr 0&0&-1&0&0&0\cr -1&0&0&0&0&0}\right)$$

For our other example, consider $SU(3)$ at level 4. There are 15 primaries,
and the vacuum is $0=(4,0,0)$. The matrix
$T$ has order $N=21$. The sign $s(0)$ is +1, while $\eps_{{1\over 2}}(0)=+1$
and $\si_{{1\over 2}}(0)=(0,2,2)$.

More generally when $k$ is even, the $T$ matrix for  $SU(3)$ will have
order $3(k+3)$. The vacuum is $0=(k,0,0)$ and $\si_{{1\over
2}}(0)=(0,{k\over 2},{k\over 2})$. As mentioned in [17], (11) implies
the following fusion product:
$$(0,{k\over 2},{k\over 2})\stimes(0,{k\over 2},{k\over
2})=(k,0,0),(k-2,1,1), \ldots, (0,{k\over 2},{k\over 2})$$
as can be verified explicitly by [18].

\bigskip\noindent{{\bf 5.\ The derivation}}\medskip

We conclude this paper with the derivation of equations (10). This is similar
to some calculations in [17].

Note that $T^{{N+1\over 2}}=s\,\sqrt{T}$, as can be seen by raising both sides
to the $N$th power.
Define $T_h:=G_{{1\over 2}}\,T^4 \,G_{{1\over 2}}^{-1}$. We will eventually
show that $T_h=T$. Hit the identity $C=(TS)^3$ with the Galois automorphism
$\si_{{2}}$. The integral matrix $C$ is unchanged, and $T$, being
a diagonal matrix with roots of 1 down the diagonal, gets sent to
$T^{{2}}$. Using (6b), we get
$$\eqalign{C=&\,T^2\,G_2\,S\,T^2\,S\,G_{{1\over 2}}\,T^2\,G_2\,S=G_2\,
T_h^{{N+1\over 2}}\,S\,T^2\,S\,T_h^{{N+1\over 2}}\,S\ ,\cr
{\rm i.e.}&\qquad 
G_{{1\over 2}}=G_2^{-1}= C\,T_h^{{N+1\over
2}}\,S\,T^{2}\,S\,T_h^{{N+1\over 2}}\,S\ .\cr}$$

Calling $D_a$ the diagonal matrix with entries $S_{ad}/S_{0d}$, equation (9)
becomes the matrix equation
$$\eqalignno{Y_a=&\,(s\,T^{{N+1\over 2}}\,S\,T^2\,S\,T^{{N+1\over 2}}\,s)\,D_a\,
(s\,T^{*{N+1\over 2}}\,S\,T^{*2}\,S\,T^{*{N+1\over 2}}\,s)&(15)\cr
=&\,
C\,s\,T^{{N+1\over 2}}\,T_h^{*{N+1\over 2}}\,G_{{1\over 2}}\,S\,D_a\,
S\,G_2\,T_h^{{N+1\over 2}}\,T^{*{N+1\over 2}}\,s&\cr}$$
where we repeatedly commute diagonal matrices.
Putting $\alpha=T_{00}T^*_{h\,00}$, the integrality (5a) of ${\cal Z}(a,b)=
Y_{a0}^b$ (5a) tells us
$$\alpha^{{N+1\over 2}}\,N_{a,\si 0}^{\si b}\,T_{h\,bb}^{{N+1\over 2}}\,
T^{*{N+1\over 2}}_{bb}\in\Z$$
for all $a,b$, where we write $\si$ for $\si_{{1\over 2}}$. Choosing any
$a$ in the fusion product of the primary field $\si 0$ with the primary
$\si b$, we get then that $\alpha\,T_{h\,bb}\,T^*_{bb}\in{\Bbb Q}$ for all
$b$. But it also must be an $N$th root of 1, and $N$ is odd, so in
fact $\alpha\,T_{h\,bb}\,T^*_{bb}=1$, i.e.\ 
$T=\alpha\,T_h$. Substituting this into (15) gives us (10a), as well as
$P=\alpha\,s\,G_{{1\over 2}}\,S\,s$. Using (6b), we find $P^2=\alpha^2\,C$,
i.e.\ $\alpha^2=1$. Hence $\alpha$, which is an $N$th root of 1 for
odd $N$, equals 1 and $P$ is given by (10b).

\bigskip\noindent{{\bf Acknowledgements}}\medskip

I thank J\"urgen Fuchs, Matthias Gaberdiel, Christoph Schweigert,
and Mark Walton for  helpful comments.
This research was supported  by NSERC.

\vfill\eject
\noindent{{\bf References}}\medskip

\item{[1]} G.\ Pradisi,   Nuovo Cimento
Soc. Ital. Fis.  {\bf B112} (1997) 467.

\item{[2]} A.\ Sagnotti and Y.\ S.\ Stanev, Nucl.\ Phys.\ Proc.\
Suppl.\ {\bf 55B} (1997) 200.

\item{[3]} J.\ Fuchs and C.\ Schweigert, Nucl.\ Phys.\ {\bf B520}
(1998)  99.

\item{[4]} Ph.\ Ruelle, ``Symmetric boundary conditions in boundary
critical phenomena'', hep-th/9904100.

\item{[5]} R.\ E.\ Behrend, P.\ A.\ Pearce, V.\ B.\ Petkova and
J.-B.\ Zuber, ``Boundary conditions in rational conformal field
theories'', hep-th/9908034.

\item{[6]} G.\ Felder, J.\ Fr\"ohlich, J.\ Fuchs and C.\ Schweigert,
``Conformal boundary conditions and three-dimensional topological
field theory'', hep-th/9909140.

\item{[7]} J.\ Cardy, Nucl.\ Phys.\ {\bf B324} (1989) 581.

\item{[8]} G.\ Pradisi, A.\ Sagnotti and Y.\ S.\ Stanev, Phys.\
Lett.\ {\bf B354} (1995) 279.

\item{[9]} L.\ Borisov, M.B.\ Halpern and C.\ Schweigert, Int.\ J.\
  Mod.\ Phys.\ {\bf A13} (1998) 125.

\item{[10]} P.\ Bantay, Phys.\ Lett.\ {\bf B419} (1998) 175.

\item{[11]} P.\ Bantay, ``Permutation orbifolds'', hep-th/9910079.

\item{[12]} G.\ Moore and N.\ Seiberg, Commun.\ Math.\ Phys.\ {\bf 123} (1989)
177.

\item{[13]} G.\ Anderson and G.\ Moore, Commun.\ Math.\ Phys.\ {\bf 117} (1988)
441.

\item{[14]} {P.\ Bantay}, Phys.\ Lett.\ {\bf B394} (1997) 87.

\item{[15]} {A.\ Coste and T.\ Gannon},
  Phys.\ Lett.\ {\bf B323} (1994) 316.

\item{[16]} L.R.\ Huiszoon, A.N.\ Schellekens and N.\ Sousa, ``Klein
  bottles and simple currents'', hep-th/9909114.

\item{[17]} A.\ Coste and T.\ Gannon, ``Congruence subgroups and RCFT'',
math.QA/9909080.

\item{[18]} L.\ B\'egin, P.\ Mathieu and M.A.\ Walton, Mod.\ Phys.\
Lett.\ {\bf A7} (1992) 3255.

\end